\documentclass[12pt]{article}
\usepackage{amssymb}

\def\sech{\mathop{\rm sech}\nolimits}
\def\csch{\mathop{\rm csch}\nolimits}
\def\deg{\mathop{\rm deg}\nolimits}
\def\case#1#2{{\textstyle{#1\over #2}}}

\title{
More on an exactly solvable position-dependent mass Schr\"odinger equation in two
dimensions: Algebraic approach and extensions to three dimensions}
\author{C. Quesne\\ 
{\sl \small Physique Nucl\'eaire Th\'eorique et Physique
Math\'ematique,  Universit\'e Libre de Bruxelles,} \\ 
{\sl \small Campus de la Plaine CP229, Boulevard~du Triomphe, B-1050 Brussels,
Belgium}\\
{\small E-mail: cquesne@ulb.ac.be}}
\date{ }
\begin{document}
\baselineskip=20pt plus 1pt minus 1pt
%%%%%%%%%%%%%%%%%%%%%%%%%%%%%%%%%%%%%%%%%%%%%%%%%%%%%%%%%%
\maketitle

\begin{abstract}

\end{abstract}
An exactly solvable position-dependent mass Schr\"odinger equation in two dimensions, depicting a
particle moving in a semi-infinite layer, is re-examined in the light of recent theories describing
superintegrable two-dimensional systems with integrals of motion that are quadratic functions of the
momenta. To get the energy spectrum a quadratic algebra approach  is used together with a
realization in terms of deformed parafermionic oscillator operators. In this process, the importance of
supplementing algebraic considerations with a proper treatment of boundary conditions for selecting
physical wavefunctions is stressed. Some new results for matrix elements are derived. Finally, the
two-dimensional model is extended to two integrable and exactly solvable (but not superintegrable)
models in three dimensions, depicting a particle in a semi-infinite parallelepipedal or cylindrical
channel, respectively.
\vspace{0.5cm}

\noindent
{\sl PACS}: 03.65.-w

\noindent
{\sl Keywords}: Schr\"odinger equation; Position-dependent mass; Quadratic algebra

\newpage
%
%======================================================================
%
\section{Introduction}

Quantum mechanical systems with a position-dependent (effective) mass (PDM) have attracted a
lot of attention and inspired intense research activites during recent years. They are indeed very
useful in the study of many physical problems, such as electronic properties of
semiconductors~\cite{bastard} and quantum dots~\cite{serra}, nuclei~\cite{ring}, quantum
liquids~\cite{arias}, $^3$He clusters~\cite{barranco}, metal clusters~\cite{puente}, etc.\par
%
%------------------------------------------------------------------------------------------------
%
Looking for exact solutions of the Schr\"odinger equation with a PDM has become an interesting
research topic because such solutions may provide a conceptual understanding of some physical
phenomena, as well as a testing ground for some approximation schemes. Although mostly
one-dimensional equations have been considered up to now, several works have recently paid
attention to $d$-dimensional problems~\cite{chen, dong, cq06, mustafa, ju, gonul}.\par
%
%---------------------------------------------------------------------------------------------
%
In \cite{cq06} (henceforth referred to as I and whose equations will be quoted by their number
preceded by I), we have analyzed the problem of $d$-dimensional PDM Schr\"odinger equations in
the framework of first-order intertwining operators and shown that with a pair $(H, H_1)$ of
intertwined Hamiltonians we can associate another pair $(R, R_1)$ of second-order partial
differential operators related to the same intertwining operator and such that $H$ (resp.\ $H_1$)
commutes with $R$ (resp.\ $R_1$). In the context of supersymmetric quantum mechanics
(SUSYQM) based on an sl(1/1) superalgebra, $R$ and $R_1$ can be interpreted as SUSY partners,
while $H$ and $H_1$ are related to the Casimir operator of a larger gl(1/1) superalgebra.\par
%
%-----------------------------------------------------------------------------------------
% 
In the same work, we have also applied our general theory to an explicit example, depicting a
particle moving in a two-dimensional semi-infinite layer. This model may be of interest in the study
of quantum wires with an abrupt termination in an environment that can be modelled by a
dependence of the carrier effective mass on the position. It illustrates the influence of a uniformity
breaking in a quantum channel on the production of bound states, as it was previously observed in
the case of a quantum dot or a bend~\cite{olendski}.\par
%
%------------------------------------------------------------------------------------------
%
{}From a theoretical viewpoint, our model has proved interesting too because it is solvable in two
different ways: by separation of variables in the corresponding Schr\"odinger equation or employing
SUSYQM and shape-invariance techniques~\cite{cooper}. The former method relies upon the
existence of an integral of motion $L$, while, as above-mentioned, the latter is based on the use of
$R$. In other words, the three second-order partial differential operators $H$, $L$ and $R$ form a
set of algebraically independent integrals of motion, which means that the system is
superintegrable.\par
%
%------------------------------------------------------------------------------------------
%
Let us recall that in classical mechanics~\cite{goldstein}, an integrable system on a
$d$-dimensional manifold is a system which has $d$ functionally independent (globally defined)
integrals of motion in involution (including the Hamiltonian). Any system with more that $d$
functionally independent integrals of motion is called superintegrable. It is maximally superintegrable
if it admits the maximum number $2d-1$ of integrals of motion. The latter form a complete set so
that any other integral of motion can be expressed in terms of them. In particular, the Poisson
bracket of any two basic integrals, being again a constant of motion, can be written as a (in
general) nonlinear function of them. Such results can be extended to quantum
mechanics~\cite{dirac}, so that for quantum counterparts of maximally superintegrable systems we
get (in general) nonlinear associative algebras of algebraically independent observables, all of them
commuting with $H$.\par
%
%------------------------------------------------------------------------------------
% 
The simplest case corresponds to the class of two-dimensional superintegrable systems with
integrals of motion that are linear and quadratic functions of the momenta. The study and
classification of such systems, dating back to the 19th century and revived in the
1960ties~\cite{fris}, have recently been the subject of intense research activites and substantial
progress has been made in this area (see \cite{hietarinta, granovskii92a, granovskii92b, bonatsos,
daska01, daska06, letourneau, ranada, tempesta, kalnins97, kalnins05} and references quoted
therein). In particular, it has been shown that their integrals of motion generate a quadratic Poisson
algebra (in the classical case) or a quadratic associative algebra (in the quantum one) with a Casimir
of sixth degree in the momenta and the general form of these algebras has been
uncovered~\cite{daska01, kalnins05}. Algebras of this kind have many similarities to the quadratic
Racah algebra QR(3) (a special case of the quadratic Askey-Wilson algebra
QAW(3))~\cite{granovskii92a}. They actually coincide with QR(3) whenever one of their
parameters vanishes. The eigenvalues and eigenfunctions of the superintegrable system
Hamiltonian can be found from the finite-dimensional irreducible representations of these algebras.
The latter can be determined by a ladder-operator method~\cite{granovskii92a, granovskii92b} or
through a realization~\cite{bonatsos, daska01} in terms of (generalized) deformed parafermionic
operators~\cite{cq94}, which are a finite-dimensional version of deformed oscillator
operators~\cite{daska91}.\par
%
%-----------------------------------------------------------------------------------------------------
%
Since our two-dimensional PDM model belongs to this class of superintegrable systems, it is
interesting to analyze it in the light of such topical and innovative theories. This is one of the
purposes of the present paper, which will therefore provide us with a third method for solving the
PDM Schr\"odinger equation. In such a process, we will insist on the necessity of supplementing
algebraic calculations with a proper treatment of the wavefunction boundary conditions imposed by
the physics of the problem --- a point that is not always highlighted enough.\par
%
%-------------------------------------------------------------------------------------------
%
The other purpose of the present paper is to free ourselves from the restriction to a
two-dimensional model. We actually plan to show that an abrupt termination of a quantum channel
can also be mimicked by some three-dimensional exactly solvable models.\par
%
%-------------------------------------------------------------------------------------------
%
This paper is organized as follows. In Section 2, the two-dimensional PDM model of I is briefly
reviewed and some important comments on its mathematical structure are made in conjunction
with the physics of the problem. The quadratic algebra approach to such a model is then detailed in
Section 3. Two three-dimensional extensions of the model are presented in Section 4. Finally,
Section 5 contains the conclusion.\par
%
%==================================================
%
\section{Exactly solvable and superintegrable PDM model in a two-dimensional semi-infinite layer}

In I we considered a particle moving in a two-dimensional semi-infinite layer of width $\pi/q$, parallel
to the $x$-axis and with impenetrable barriers at the boundaries. The variables $x$, $y$ vary in the
domain
\begin{equation}
  D: \qquad 0 < x < \infty, \qquad - \frac{\pi}{2q} < y < \frac{\pi}{2q},  \label{eq:D2} 
\end{equation}
and the wavefunctions have to satisfy the conditions
\begin{equation}
  \psi(0,y) = 0, \qquad \psi\left(x, \pm \frac{\pi}{2q}\right) = 0.  \label{eq:boundary2}
\end{equation}
The mass of the particle is $m(x) = m_0 M(x)$, where the dimensionless function $M(x)$ is given
by
\begin{equation}
  M(x) = \sech^2 qx.  \label{eq:mass}
\end{equation}
In units wherein $\hbar = 2 m_0 = 1$, the Hamiltonian of the model can be written as 
\begin{equation}
  H^{(k)} = - \partial_x \frac{1}{M(x)} \partial_x - \partial_y \frac{1}{M(x)} \partial_y +
  V^{(k)}_{\rm eff}(x),  \label{eq:H2}
\end{equation}
where
\begin{equation}
  V^{(k)}_{\rm eff}(x) = - q^2 \cosh^2 qx + q^2 k(k-1) \csch^2 qx  \label{eq:Veff}
\end{equation}
is an effective potential including terms depending on the ambiguity parameters (see Eq.~(I2.3)). In
(\ref{eq:Veff}), the constant $k$ is assumed positive and we have set an irrelevant additive
constant $v_0$ to zero.\par
%
%-----------------------------------------------------------------------------------
%
Both the operators
\begin{equation}
  L = - \partial_y^2
\end{equation}
and
\begin{eqnarray}
  R^{(k)} & = & \eta^{(k)\dagger} \eta^{(k)} \nonumber \\ 
  & = & - \cosh^2 qx \sin^2 qy\, \partial^2_x + 2 \sinh qx \cosh qx \sin qy
       \cos qy\, \partial^2_{xy} - \sinh^2 qx \cos^2 qy\, \partial^2_y \nonumber \\
  && \mbox{} + q \sinh qx \cosh qx (1 - 4 \sin^2 qy) \partial_x
       + q (1 + 4 \sinh^2 qx) \sin qy \cos qy \partial_y \nonumber \\
  && \mbox{} + q^2 (\sinh^2 qx - \sin^2 qy - 3 \sinh^2 qx \sin^2 qy) - q^2 k (1 + 
       \csch^2 qx \sin^2 qy) \nonumber \\
  && \mbox{} + q^2 k^2 \csch^2 qx \sin^2 qy,  
\end{eqnarray}
where
\begin{eqnarray}
  \eta^{(k)\dagger} & = & - \cosh qx \sin qy \,\partial_x + \sinh qx \cos qy\, \partial_y - q 
       \sinh qx \sin qy \nonumber \\
  && \mbox{} - qk \csch qx \sin qy, \\  
  \eta^{(k)} & = & \cosh qx \sin qy \,\partial_x - \sinh qx \cos qy\, \partial_y + q 
       \sinh qx \sin qy \nonumber \\
  && \mbox{} - qk \csch qx \sin qy, 
\end{eqnarray}
commute with $H^{(k)}$, although not with one another. Hence one may diagonalize either
$H^{(k)}$ and $L$ or $H^{(k)}$ and $R^{(k)}$ simultaneously. This leads to two alternative bases
for the Hamiltonian eigenfunctions, corresponding to the eigenvalues
\begin{equation}
  E^{(k)}_N = q^2 (N+2) (N+2k+1), \qquad N=0, 1, 2, \ldots,  \label{eq:E}
\end{equation}
with degeneracies
\begin{equation}
  \deg(N) = \left[\frac{N}{2}\right] + 1,  \label{eq:deg}
\end{equation}
where $[N/2]$ stands for the integer part of $N/2$.\par
%
%---------------------------------------------------------------------------------
%
The first basis is obtained by separating the variables $x$, $y$ in the PDM Schr\"odinger equation
and its members, associated with the eigenvalues $(l+1)^2 q^2$ of $L$, read
\begin{equation}
  \psi^{(k)}_{n,l}(x,y) = \phi^{(k)}_{n,l}(x) \chi_l(y), \qquad  n, l = 0, 1, 2, \ldots, \label{eq:psi} 
\end{equation}
with $N = 2n+l$,
\begin{equation}
  \phi^{(k)}_{n,l} = {\cal N}^{(k)}_{n,l} (\tanh qx)^k (\sech qx)^{l+2}
       P^{\left(k-\case{1}{2}, l+1\right)}_n(1 - 2 \tanh^2 qx),
\end{equation}
\begin{equation}
  \chi_l(y) = \left\{\begin{array}{ll}
      \sqrt{\frac{2q}{\pi}} \cos[(l+1)qy] & \qquad {\rm for\ } l = 0, 2, 4, \ldots, \\[0.2cm]
      \sqrt{\frac{2q}{\pi}} \sin[(l+1)qy] & \qquad {\rm for\ } l = 1, 3, 5, \ldots,
  \end{array} \right.  \label{eq:chi}
\end{equation}
and ${\cal N}^{(k)}_{n,l}$ given in Eq.~(I3.18).\par
%
%-------------------------------------------------------------------------------------
%
The second basis, resulting from the intertwining relation
\begin{equation}
  \eta^{(k)} H^{(k)} = H_1^{(k)} \eta^{(k)}, \qquad H_1^{(k)} = H^{(k+1)} + 2 q^2 k,
\end{equation}
and its Hermitian conjugate, can be built by successive applications of operators of type $\eta^{(k)
\dagger}$,
\begin{equation}
  \Psi^{(k)}_{N,N_0}(x,y) = \bar{\cal N}^{(k)}_{N,N_0} \eta^{(k)\dagger}
  \eta^{(k+1)\dagger} \cdots \eta^{(k+\nu-1)\dagger} \Psi^{(k+\nu)}_{N_0,N_0}(x,y),
  \label{eq:Psi} 
\end{equation}
on functions $\Psi^{(k+\nu)}_{N_0,N_0}(x,y)$, annihilated by $\eta^{(k+\nu)}$ and given in
Eqs.~(I3.28), (I3.32) and (I3.34). In (\ref{eq:Psi}), $N_0$ runs over 0, 2, 4,\ldots, $N$ or $N-1$,
according to whether $N$ is even or odd, while $\nu$, defined by $\nu = N - N_0$, determines the
$R^{(k)}$ eigenvalue
\begin{equation}
  r^{(k)}_{\nu} = q^2 \nu (\nu + 2k), \qquad \nu=0, 1, 2, \ldots.  \label{eq:r}
\end{equation}
Although an explicit expression of the normalization coefficient $\bar{\cal N}^{(k)}_{N,N_0}$ is
easily obtained (see Eq.~(I3.41)), this is not the case for $\Psi^{(k)}_{N,N_0}(x,y)$ (except for
some low values of $N$ and $N_0$), nor for the expansion of $\Psi^{(k)}_{N,N_0}(x,y)$ into the
first basis eigenfunctions $\psi^{(k)}_{n,l}(x,y)$, which is given by rather awkward formulas (see
Eqs.~(I3.46), (I3.51), (I3.55) and (I3.56)).\par
%
%--------------------------------------------------------------------------------------------------
%
Before proceeding to a quadratic algebra approach to the problem in Section 3, it is worth
making a few valuable observations.\par
%
%------------------------------------------------------------------------------------------------
%
Mathematically speaking, the separable Schr\"odinger equation of our model admits four linearly
independent solutions obtained by combining the two independent solutions of the second-order
differential equation in $x$ with those of the second-order differential equation in $y$. Among
those four functions, only the combination $\psi^{(k)}_{n,l}(x,y)$, considered in (\ref{eq:psi}),
satisfies all the boundary conditions and is normalizable on $D$. It is indeed clear that the alternative
solution to the differential equation in $x$ is not normalizable, while that to the differential
equation in $y$,
\begin{equation}
  \bar{\chi}_l(y) \propto \left\{\begin{array}{ll}
      \sin[(l+1)qy] & \qquad {\rm for\ } l = 0, 2, 4, \ldots, \\[0.2cm]
      \cos[(l+1)qy] & \qquad {\rm for\ } l = -1, 1, 3, 5, \ldots,
  \end{array} \right.  \label{eq:chi-bar}  
\end{equation}
violates the second condition in Eq.~(\ref{eq:boundary2}). Hence the three remaining combinations
provide unphysical functions.\par
%
%-------------------------------------------------------------------------------------------
%
Some mathematical considerations might also lead to another choice than $L$ and $R^{(k)}$ for
the basic integrals of motion complementing $H^{(k)}$. First of all, instead of $L$, one might
select the operator $p_y = - {\rm i} \partial_y$, which obviously satisfies the condition
$[H^{(k)}, p_y] = 0$. This would result in a linear and a quadratic (in the momenta) integrals of
motion, generating a much simpler quadratic algebra than that to be considered in Section 3. It
should be realized, however, that the eigenfunctions $e^{{\rm i}my}$ ($m \in \mathbf{Z}$) of
$p_y$, being linear combinations of the physical and unphysical functions (\ref{eq:chi}) and
(\ref{eq:chi-bar}), are useless from a physical viewpoint. We are therefore forced to consider the
second-order operator $L$ instead of $p_y$.\par
%
%-----------------------------------------------------------------------------------------------
%
{}Furthermore, it is straightforward to see that another pair of first-order differential operators
\begin{eqnarray}
  \bar{\eta}^{(k)\dagger} & = & - \cosh qx \cos qy \,\partial_x - \sinh qx \sin qy\, \partial_y - q 
       \sinh qx \cos qy \nonumber \\
  && \mbox{} - qk \csch qx \cos qy,  \label{eq:eta-bar-plus} \\  
  \bar{\eta}^{(k)} & = & \cosh qx \cos qy \,\partial_x + \sinh qx \sin qy\, \partial_y + q 
       \sinh qx \cos qy \nonumber \\
  && \mbox{} - qk \csch qx \cos qy,  \label{eq:eta-bar} 
\end{eqnarray}
intertwines with $H^{(k)}$ and $H_1{(k)}$, i.e., satisfies the relation
\begin{equation}
  \bar{\eta}^{(k)} H^{(k)} = H_1^{(k)} \bar{\eta}^{(k)}, \qquad H_1^{(k)} = H^{(k+1)} + 2 q^2 k,
  \label{eq:inter-bar}
\end{equation}
and its Hermitian conjugate. Such operators correspond to the choice $a = c = g = 0$, $b = d = 1$
in Eq.~(I2.29).\par
%
%--------------------------------------------------------------------------------
%
As a consequence of (\ref{eq:inter-bar}), the operator
\begin{eqnarray}
  \bar{R}^{(k)} & = & \bar{\eta}^{(k)\dagger} \bar{\eta}^{(k)} \nonumber \\ 
  & = & - \cosh^2 qx \cos^2 qy\, \partial^2_x - 2 \sinh qx \cosh qx \sin qy
       \cos qy\, \partial^2_{xy} - \sinh^2 qx \sin^2 qy\, \partial^2_y \nonumber \\
  && \mbox{} + q \sinh qx \cosh qx (1 - 4 \cos^2 qy) \partial_x
       - q (1 + 4 \sinh^2 qx) \sin qy \cos qy \partial_y \nonumber \\
  && \mbox{} + q^2 (\sinh^2 qx - \cos^2 qy - 3 \sinh^2 qx \cos^2 qy) - q^2 k (1 + 
       \csch^2 qx \cos^2 qy) \nonumber \\
  && \mbox{} + q^2 k^2 \csch^2 qx \cos^2 qy,  
\end{eqnarray}
commutes with $H^{(k)}$ and is therefore another integral of motion. It can of course be
expressed in terms of $H^{(k)}$, $L$ and $R^{(k)}$, as it can be checked that
\begin{equation}
  H^{(k)} = L + R^{(k)} + \bar{R}^{(k)} + 2 q^2 k.
\end{equation}
However, we have now at our disposal three (dependent) integrals of motion $L$, $R^{(k)}$ and
$\bar{R}^{(k)}$ in addition to $H^{(k)}$, so that we may ask the following question: what is the
best choice for the basic integrals of motion from a physical viewpoint?\par
%
%-----------------------------------------------------------------------------------
%
This problem is easily settled by noting that the zero modes of $\bar{\eta}^{(k)}$,
\begin{equation}
  \bar{\omega}^{(k)}_s(x,y) = (\tanh qx)^k (\sech qx)^{s+1} (\sin qy)^s,
\end{equation}
violate the second condition in Eq.~(\ref{eq:boundary2}) for any real value of $s$ and therefore
lead to unphysical functions. This contrasts with what happens for the zero modes
$\omega^{(k)}_s(x,y)$ of $\eta^{(k)}$, given in (I3.28), which are physical functions for $s > 0$
and can therefore be used to build the functions $\Psi^{(k)}_{N,N_0}(x,y)$ considered in
(\ref{eq:Psi}), as it was shown in (I3.32). We conclude that the physics of the model imposes
the choice of $L$ and $R^{(k)}$ as basic integrals of motion.\par
%
%====================================================
%
\section{Quadratic algebra approach to the PDM model in a two-dimensional semi-infinite layer}

\setcounter{equation}{0}

\subsection{Quadratic associative algebra and its classical limit}

It has been shown~\cite{daska01, kalnins05} that for any two-dimensional quantum superintegrable
system with integrals of motion $A$, $B$, which are second-order differential operators, one can
construct a quadratic associative algebra generated by $A$, $B$, and their commutator $C$. This
operator is not independent of $A$, $B$, but since it is a third-order differential operator, it cannot
be written as a polynomial function of them. The general form of the quadratic algebra
commutation relations is
\begin{eqnarray}
  [A, B] & = & C,  \label{eq:C1} \\{}
  [A, C] & = & \alpha A^2 + \gamma \{A, B\} + \delta A + \epsilon B + \zeta,  \label{eq:C2} \\{}
  [B, C] & = & a A^2 - \gamma B^2 - \alpha \{A, B\} + d A - \delta B + z.  \label{eq:C3}
\end{eqnarray}
Here $\{A, B\} \equiv AB + BA$,
\begin{eqnarray}
  \delta & = & \delta(H) = \delta_0 + \delta_1 H, \qquad \epsilon = \epsilon(H) = \epsilon_0 +
        \epsilon_1 H, \qquad \zeta = \zeta(H) = \zeta_0 + \zeta_1 H + \zeta_2 H^2, \nonumber
        \\
  d & = & d(H) = d_0 + d_1 H, \qquad z = z(H) = z_0 + z_1 H + z_2 H^2,   
\end{eqnarray} 
and $\alpha$, $\gamma$, $a$, $\delta_i$, $\epsilon_i$, $\zeta_i$, $d_i$, $z_i$ are some
constants. Note that it is the Jacobi identity $[A, [B, C]] = [B, [A, C]]$ that imposes some
relations between coefficients in (\ref{eq:C2}) and (\ref{eq:C3}).\par
%
%-----------------------------------------------------------------------------------------------
% 
Such a quadratic algebra closes at level 6~\cite{kalnins05} or, in other words, it has a Casimir
operator which is a sixth-order differential operator~\cite{daska01},
\begin{eqnarray}
  K & = & C^2 + \case{2}{3} a A^3 - \case{1}{3} \alpha \{A, A, B\} - \case{1}{3} \gamma \{A, B,
        B\} + \left(\case{2}{3} \alpha^2 + d + \case{2}{3} a \gamma\right) A^2 \nonumber \\
  && \mbox{} + \left(\case{1}{3} \alpha \gamma - \delta\right) \{A, B\} + \left(\case{2}{3}
        \gamma^2 - \epsilon\right) B^2 + \left(\case{2}{3} \alpha \delta + \case{1}{3} a \epsilon
        + \case{1}{3} d \gamma + 2z\right) A \nonumber \\
  && \mbox{} + \left(- \case{1}{3} \alpha \epsilon + \case{2}{3} \gamma \delta - 2 \zeta\right)
        B + \case{1}{3} \gamma z - \case{1}{3} \alpha \zeta \nonumber \\
  & = & k_0 + k_1 H + k_2 H^2 + k_3 H^3,  \label{eq:K}  
\end{eqnarray}
where $k_i$ are some constants and $\{A, B, C\} \equiv ABC + ACB + BAC + BCA + CAB +
CBA$.\par
%
%-------------------------------------------------------------------------------------------
%
{}For our two-dimensional PDM model, described by the Hamiltonian defined in Eqs.~(\ref{eq:mass})
-- (\ref{eq:Veff}), we shall take
\begin{equation}
  A = R, \qquad B = L,  \label{eq:A-B}
\end{equation}
where, for simplicity's sake, we dropped the superscript $(k)$ because no confusion can arise
outside the SUSYQM context.\par
%
%--------------------------------------------------------------------------------------------
%
To determine their commutation relations, it is worth noting first that their building blocks, the
first-order differential operators $\partial_y$, $\eta^{\dagger}$ and $\eta$, generate another
quadratic algebra together with the other set of intertwining operators $\bar{\eta}^{\dagger}$,
$\bar{\eta}$, given in (\ref{eq:eta-bar-plus}) and (\ref{eq:eta-bar}). Their commutation relations
are indeed easily obtained as
\begin{eqnarray}
  [\partial_y, \eta] & = & q \bar{\eta}, \qquad [\partial_y, \bar{\eta}] = - q \eta, \qquad 
       [\eta, \bar{\eta}] = q \partial_y, \\{}
  [\eta, \eta^{\dagger}] & = & 2 q^2 k (1 + \xi^2), \qquad  [\bar{\eta}, \bar{\eta}^{\dagger}] =
       2 q^2 k (1 + \bar{\xi}^2), \qquad [\eta, \bar{\eta}^{\dagger}] = - q  \partial_y + 2 q^2 k  \xi
       \bar{\xi},  \label{eq:com-eta}
\end{eqnarray}
and their Hermitian conjugates. In (\ref{eq:com-eta}), we have defined
\begin{equation}
  \xi = - (2qk)^{-1} (\eta + \eta^{\dagger}) = \csch qx \sin qy, \qquad \bar{\xi} = - (2qk)^{-1}
  (\bar{\eta} + \bar{\eta}^{\dagger}) = \csch qx \cos qy. 
\end{equation}
Interestingly, $\partial_y$, $\eta$ and $\bar{\eta}$ (as well as $\partial_y$, $\eta^{\dagger}$
and $\bar{\eta}^{\dagger}$) close an sl(2) subalgebra.\par
%
%-----------------------------------------------------------------------------------
%
{}From these results, it is now straightforward to show that the operator $C$ in (\ref{eq:C1}) is
given by
\begin{equation}
  C = q \{\partial_y, \eta^{\dagger} \bar{\eta} + \bar{\eta}^{\dagger} \eta\}
\end{equation}
and that the coefficients in (\ref{eq:C2}) and (\ref{eq:C3}) are
\begin{eqnarray}
  \alpha & = & \gamma = 8 q^2, \qquad \delta = 8 q^2 [q^2 (2k-1) - H], \qquad \epsilon = 16 q^4
       (k-1)(k+1), \nonumber \\
  \zeta & = & 8 q^4 (k-1) (2 q^2 k - H), \qquad a = 0, \qquad d = 16 q^4, \qquad z = 8 q^4 (2
       q^2 k - H).  \label{eq:parameters}
\end{eqnarray}
On inserting the latter in (\ref{eq:K}), we obtain for the value of the Casimir operator
\begin{equation}
  K = - 4 q^4 [2q^2 (7k-6) - 3H] (2q^2 k - H).
\end{equation}
It is worth noting that since $a=0$ in (\ref{eq:C3}), we actually have here an example of quadratic
Racah algebra QR(3)~\cite{granovskii92a}.\par
%
%---------------------------------------------------------------------------------------------
%
Before proceeding to a study of its finite-dimensional irreducible representations in Section 3.2, it is
interesting to consider its classical limit. For such a purpose, since we have adopted units wherein
$\hbar = 2 m_0 = 1$, we have first to make a change of variables and of parameters restoring a
dependence on $\hbar$ (but keeping $2 m_0 = 1$ for simplicity's sake) before letting $\hbar$ go
to zero.\par
%
%-----------------------------------------------------------------------------------------------
%
An appropriate transformation is
\begin{equation}
  X = \hbar x, \qquad Y = \hbar y, \qquad P_X = - {\rm i} \hbar \partial_X, \qquad P_Y = - {\rm i}
  \hbar \partial_Y, \qquad Q = \frac{q}{\hbar}, \qquad K = \hbar k. 
\end{equation}
On performing it on the Hamiltonian given in Eqs.~(\ref{eq:mass}) -- (\ref{eq:Veff}), we obtain
\begin{equation}
  H = - \hbar^2 (\partial_X \cosh^2 QX \partial_X + \partial_Y \cosh^2 QX \partial_Y) - \hbar^2
  Q^2 \cosh^2 QX + Q^2 K(K - \hbar) \csch^2 QX, 
\end{equation}
yielding the classical Hamiltonian
\begin{equation}
  H_{\rm c} = \lim_{\hbar \to 0} H = \cosh^2 QX (P_X^2 + P_Y^2) + Q^2 K^2 \csch^2 QX.
\end{equation}
A similar procedure applied to the intertwining operators leads to
\begin{eqnarray}
  \eta_{\rm c} & = & \lim_{\hbar \to 0} \eta \nonumber \\
  & = & {\rm i} \cosh QX \sin QY P_X - {\rm i} \sinh QX \cos QY P_Y - QK \csch QX \sin QY,  \\
  \bar{\eta}_{\rm c} & = & \lim_{\hbar \to 0} \bar{\eta} \nonumber \\
  & = & {\rm i} \cosh QX \cos QY P_X + {\rm i} \sinh QX \sin QY P_Y - QK \csch QX \cos QY, 
\end{eqnarray}
together with $\eta^*_{\rm c} = \lim_{\hbar \to 0} \eta^{\dagger}$ and $\bar{\eta}^*_{\rm c} =
\lim_{\hbar \to 0} \bar{\eta}^{\dagger}$, while the operators quadratic in the momenta give
rise to the functions
\begin{eqnarray}
  L_{\rm c} & = & \lim_{\hbar \to 0} L = P_Y^2, \\
  R_{\rm c} & = & \lim_{\hbar \to 0} R = \cosh^2 QX \sin^2 QY P_X^2 - 2 \sinh QX \cosh QX
       \sin QY \cos QY P_X P_Y \nonumber \\
  && \mbox{} + \sinh^2 QX \cos^2 QY P_Y^2 + Q^2 K^2 \csch^2 QX \sin^2 QY, \\
  \bar{R}_{\rm c} & = & \lim_{\hbar \to 0} \bar{R} = \cosh^2 QX \cos^2 QY P_X^2 + 2 \sinh QX
       \cosh QX \sin QY \cos QY P_X P_Y \nonumber \\
  && \mbox{} + \sinh^2 QX \sin^2 QY P_Y^2 + Q^2 K^2 \csch^2 QX \cos^2 QY, 
\end{eqnarray}   
satisfying the relation
\begin{equation}
  H_{\rm c} = L_{\rm c} + R_{\rm c} + \bar{R}_{\rm c}.
\end{equation}
\par
%
%--------------------------------------------------------------------------------------------------
%
The quadratic associative algebra (\ref{eq:C1}) -- (\ref{eq:K}) is now changed into a quadratic
Poisson algebra, whose defining relations can be determined either by taking the limit $
\lim_{\hbar \to 0} ({\rm i} \hbar)^{-1} [O, O'] = \{O_{\rm c}, O'_{\rm c}\}_{\rm P}$ or by direct
calculation of the Poisson brackets $\{O_{\rm c}, O'_{\rm c}\}_{\rm P}$:
\begin{eqnarray}
  \{A_{\rm c}, B_{\rm c}\}_{\rm P} & = & C_{\rm c}, \\
  \{A_{\rm c}, C_{\rm c}\}_{\rm P} & = & \alpha_{\rm c} A_{\rm c}^2 + 2 \gamma_{\rm c} 
        A_{\rm c} B_{\rm c} + \delta_{\rm c} A_{\rm c} + \epsilon_{\rm c} B_{\rm c} +
        \zeta_{\rm c}, \\
  \{B_{\rm c}, C_{\rm c}\}_{\rm P} & = & a_{\rm c} A_{\rm c}^2 - \gamma_{\rm c} 
        B_{\rm c}^2 - 2 \alpha_{\rm c} A_{\rm c} B_{\rm c} + d_{\rm c} A_{\rm c} - \delta_{\rm c}
        B_{\rm c} + z_{\rm c}. 
\end{eqnarray}
Here
\begin{equation}
  C_{\rm c} = \lim_{\hbar \to 0} \frac{C}{{\rm i} \hbar} = 2Q P_Y (\eta_{\rm c}^*
  \bar{\eta}_{\rm c} + \bar{\eta}_{\rm c}^* \eta_{\rm c})
\end{equation}
and
\begin{equation}
  \alpha_{\rm c} = \gamma_{\rm c} = - 8 Q^2, \qquad \delta_{\rm c} = 8 Q^2 H_{\rm c}, \qquad
  \epsilon_{\rm c} = - 16 Q^4 K^2, \qquad \zeta_{\rm c} = a_{\rm c} = d_{\rm c} = z_{\rm c} =
  0.
\end{equation}
Such a Poisson algebra has a vanishing Casimir:
\begin{equation}
  K_{\rm c} = \lim_{\hbar \to 0} K = 0.
\end{equation}
\par
%
%++++++++++++++++++++++++++++++++++++++++++++++++++++++++++++
%
\subsection{Finite-dimensional irreducible representations of the quadratic associative algebra}

The quadratic algebra (\ref{eq:C1}) -- (\ref{eq:K}) can be realized in terms of (generalized)
deformed oscillator operators $\cal N$, $b^{\dagger}$, $b$, satisfying the
relations~\cite{daska91}
\begin{equation}
  [{\cal N}, b^{\dagger}] = b^{\dagger}, \qquad [{\cal N}, b] = - b, \qquad b^{\dagger} b = 
  \Phi({\cal N}), \qquad b b^{\dagger} = \Phi({\cal N}+1),
\end{equation}
where the structure function $\Phi(x)$ is a `well-behaved' real function such that
\begin{equation}
  \Phi(0) = 0, \qquad \Phi(x) > 0 \quad {\rm for} \quad x > 0.  \label{eq:Phi-C1}
\end{equation}
This deformed oscillator algebra has a Fock-type representation, whose basis states $|m\rangle$,
$m=0$, 1, 2,~\ldots,\footnote{We adopt here the unusual notation $|m\rangle$ in order to avoid
confusion between the number of deformed bosons and the quantum number $n$ introduced in
(\ref{eq:psi}).} fulfil the relations
\begin{equation}
\begin{array}{l}
  {\cal N} |m\rangle = m |m\rangle, \\[10pt]
  b^{\dagger} |m\rangle = \sqrt{\Phi(m+1)}\, |m+1\rangle, \qquad m = 0, 1, 2, \ldots, \\[10pt]
  b |0\rangle = 0, \\[10pt]  
  b |m\rangle = \sqrt{\Phi(m)}\, |m-1\rangle, \qquad m = 1, 2, \ldots.
\end{array}  \label{eq:Fock}
\end{equation}
\par
%
%---------------------------------------------------------------------------------
%
We shall be more specifically interested here in a subclass of deformed oscillator operators, which
have a ($p+1$)-dimensional Fock space, spanned by $|p, m\rangle \equiv |m\rangle$, $m=0$,
1,~\ldots, $p$, due to the following property
\begin{equation}
  \Phi(p+1) = 0  \label{eq:Phi-C2}
\end{equation}
of the structure function, implying that
\begin{equation}
  (b^{\dagger})^{p+1} = b^{p+1} = 0.
\end{equation}
These are so-called (generalized) deformed parafermionic oscillator operators of order
$p$~\cite{cq94}. The general form of their structure function is given by
\begin{equation}
  \Phi(x) = x (p+1-x) (a_0 + a_1 x + a_2 x^2 + \cdots + a_{p-1} x^{p-1}),
\end{equation}
where $a_0$, $a_1$,~\ldots, $a_{p-1}$ may be any real constants such that the second condition
in (\ref{eq:Phi-C1}) is satisfied for $x=1$, 2,~\ldots, $p$.\par
%
%-----------------------------------------------------------------------------------------------
% 
A realization of the quadratic algebra (\ref{eq:C1}) -- (\ref{eq:K}) in terms of deformed oscillator
operators $\cal N$, $b^{\dagger}$, $b$ reads~\cite{daska01}
\begin{eqnarray}
  A & = & A({\cal N}), \label{eq:A-para} \\
  B & = & \sigma({\cal N}) + b^{\dagger} \rho({\cal N}) + \rho({\cal N}) b,  \label{eq:B-para} 
\end{eqnarray}
where $A({\cal N})$, $\sigma({\cal N})$ and $\rho({\cal N})$ are some functions of $\cal N$,
which, in the $\gamma \ne 0$ case, are given by
\begin{eqnarray}
  A({\cal N}) & = & \frac{\gamma}{2} \left[({\cal N}+u)^2 - \frac{1}{4} -
         \frac{\epsilon}{\gamma^2}\right], \\
  \sigma({\cal N}) & = & - \frac{\alpha}{4} \left[({\cal N}+u)^2 - \frac{1}{4}\right] +
         \frac{\alpha\epsilon - \gamma\delta}{2 \gamma^2} - \frac{\alpha\epsilon^2 - 2 \gamma
         \delta\epsilon + 4 \gamma^2 \zeta}{4 \gamma^4} \frac{1}{({\cal N}+u)^2 - \frac{1}{4}},
         \label{eq:sigma} \\
  \rho^2({\cal N}) & = & \frac{1}{3 \cdot 2^{12} \gamma^8 ({\cal N}+u) ({\cal N}+u+1) [2 ({\cal
         N}+u)+1]^2},  \label{eq:rho}
\end{eqnarray}
with the structure function
\begin{eqnarray}
  \Phi(x) & = & - 3072 \gamma^6 K [2 ({\cal N}+u)-1]^2 \nonumber \\
  && \mbox{} - 48 \gamma^6 (\alpha^2 \epsilon - \alpha\gamma\delta + a\gamma\epsilon -
       d\gamma^2) [2 ({\cal N}+u)-3] [2 ({\cal N}+u)-1]^4 [2 ({\cal N}+u)+1] \nonumber \\
  && \mbox{} + \gamma^8 (3\alpha^2 + 4a\gamma) [2 ({\cal N}+u)-3]^2 [2 ({\cal N}+u)-1]^4
       [2 ({\cal N}+u)+1]^2 \nonumber \\
  && \mbox{} + 768 (\alpha\epsilon^2 - 2\gamma\delta\epsilon + 4\gamma^2 \zeta)^2
       \nonumber \\
  && \mbox{} + 32 \gamma^4 (3\alpha^2 \epsilon^2 - 6\alpha\gamma\delta\epsilon +
       2a\gamma\epsilon^2 + 2\gamma^2 \delta^2 - 4d\gamma^2 \epsilon + 8\gamma^3 z +
       4\alpha\gamma^2 \zeta) \nonumber \\
  && \quad \mbox{}\times [2 ({\cal N}+u)-1]^2 [12({\cal N}+u)^2 - 12({\cal N}+u) - 1] \nonumber
\\
  && \mbox{} - 256 \gamma^2 (3\alpha^2 \epsilon^3 - 9\alpha\gamma\delta\epsilon^2 +
       a\gamma\epsilon^3 + 6\gamma^2 \delta^2 \epsilon - 3d\gamma^2 \epsilon^2 +
       2\gamma^4 \delta^2 + 2d\gamma^4 \epsilon + 12\gamma^3 \epsilon z \nonumber \\
  && \quad \mbox{} - 4\gamma^5 z + 12\alpha\gamma^2 \epsilon\zeta - 12\gamma^3 \delta
       \zeta + 4\alpha\gamma^4 \zeta) [2 ({\cal N}+u)-1]^2.  \label{eq:Phi-gen}   
\end{eqnarray}
These functions depend upon two (so far undetermined) constants, $u$ and the eigenvalue of the
Casimir operator $K$ (which we denote by the same symbol).\par
%
%----------------------------------------------------------------------------------------------
% 
Such a realization is convenient to determine the representations of the quadratic algebra in a basis
wherein the generator $A$ is diagonal together with $K$ (or, equivalently, $H$) because the
former is already diagonal with eigenvalues $A(m)$. The ($p+1$)-dimensional representations,
associated with ($p+1$)-fold degenerate energy levels, correspond to the restriction to deformed
parafermionic operators of order $p$~\cite{daska01}. The first condition in (\ref{eq:Phi-C1}) can
then be used with Eq.~(\ref{eq:Phi-C2}) to compute $u$ and $K$ (or $E$) in terms of $p$ and the
Hamiltonian parameters. A choice is then made between the various solutions that emerge from the
calculations by imposing the second restriction in (\ref{eq:Phi-C1}) for $x=1$, 2,~\ldots, $p$.\par
%
%----------------------------------------------------------------------------------------------------------------
%
In the present case, for the set of parameters (\ref{eq:parameters}), the complicated structure
function (\ref{eq:Phi-gen}) drastically simplifies to yield the factorized expression
\begin{eqnarray}
  \Phi(x) & = & 3 \cdot 2^{30} q^{20} (2x+2u+k-1) (2x+2u+k-2) (2x+2u-k) (2x+2u-k-1)
       \nonumber \\
  && \mbox{} \times \left(2x+2u- \case{1}{2} + \Delta\right) \left(2x+2u - \case{3}{2} +
       \Delta\right) \left(2x+2u- \case{1}{2} - \Delta\right) \nonumber \\
  && \mbox{} \times \left(2x+2u- \case{3}{2} - \Delta\right),
\end{eqnarray}
where
\begin{equation}
  \Delta = \sqrt{\left(k - \frac{1}{2}\right)^2 + \frac{E}{q^2}}.
\end{equation}
Furthermore, the eigenvalues of the operator $A$ become
\begin{equation}
  A(m) = q^2 (2m+2u-k) (2m+2u+k).
\end{equation}
Since $A=R$ is a positive-definite operator, only values of $u$ such that $A(m) \ge 0$ for $m=0$,
1,~\ldots, $p$ should be retained.\par
%
%-----------------------------------------------------------------------------------------------
%
On taking this into account, the first condition in (\ref{eq:Phi-C1}) can be satisfied by choosing
either $u = k/2$ or $u = (k+1)/2$, yielding
\begin{equation}
  A(m) = 4 q^2 m(m+k)  \label{eq:A-1}
\end{equation}
or
\begin{equation}
  A(m) = 4 q^2 \left(m + \case{1}{2}\right) \left(m + k + \case{1}{2}\right),  \label{eq:A-2}
\end{equation}
respectively. For $u = k/2$, Eq.~(\ref{eq:Phi-C2}), together with the second condition in
(\ref{eq:Phi-C1}), can be fulfilled in two different ways corresponding to $\Delta = 2p + k + 1
\pm \frac{1}{2}$ or
\begin{equation}
  E = q^2 \left(2p + \case{3}{2} \pm \case{1}{2}\right) \left(2p + 2k + \case{1}{2} \pm
  \case{1}{2}\right).  \label{eq:E-1}
 \end{equation}
The resulting structure function reads
\begin{eqnarray}
  \Phi(x) & = & 3 \cdot 2^{38} q^{20} x (p+1-x) \left(x - \case{1}{2}\right) \left(p+1 \pm
       \case{1}{2} - x\right) \left(x+k - \case{1}{2}\right) (x+k-1) \nonumber \\
  && \mbox{} \times \left(x+p+k + \case{1}{4} \pm \case{1}{4}\right) \left(x+p+k -
       \case{1}{4} \pm \case{1}{4}\right).  \label{eq:Phi-1} 
\end{eqnarray}
Similarly, for $u = (k+1)/2$, we obtain
\begin{equation}
  E = q^2 \left(2p + \case{5}{2} \pm \case{1}{2}\right) \left(2p + 2k + \case{3}{2} \pm
  \case{1}{2}\right)  \label{eq:E-2}
 \end{equation}
and
\begin{eqnarray}
  \Phi(x) & = & 3 \cdot 2^{38} q^{20} x (p+1-x) \left(x + \case{1}{2}\right) \left(p+1 \pm
       \case{1}{2} - x\right) (x+k) \left(x+k - \case{1}{2}\right)  \nonumber \\
  && \mbox{} \times \left(x+p+k + \case{5}{4} \pm \case{1}{4}\right) \left(x+p+k +
       \case{3}{4} \pm \case{1}{4}\right).  \label{eq:Phi-2} 
\end{eqnarray}
\par
%
%--------------------------------------------------------------------------------------------
%
Our quadratic algebra approach has therefore provided us with a purely algebraic derivation of the
eigenvalues of $H$ and $R$ in a basis wherein they are simultaneously diagonal. It now remains to
see to which eigenvalues we can associate physical wavefunctions, i.e., normalizable functions
satisfying Eq.~(\ref{eq:boundary2}). This will imply a correspondence between $|p, m\rangle$ and
the functions $\Psi_{N, N-\nu}(x,y)$, defined in (\ref{eq:Psi}).\par
%
%--------------------------------------------------------------------------------------------
%
On comparing $A(m)$ to the known (physical) eigenvalues $r_{\nu}$ of $R$, given in (\ref{eq:r}),
we note that the first choice (\ref{eq:A-1}) for $A(m)$ corresponds to even $\nu = 2m$ (hence
to even $N$), while the second choice (\ref{eq:A-2}) is associated with odd $\nu = 2m+1$ (hence
with odd $N$). Appropriate values of $p$ leading to the level degeneracies (\ref{eq:deg}) are
therefore $p = N/2$ and $p = (N-1)/2$, respectively. With this identification, both
Eqs.~(\ref{eq:E-1}) and (\ref{eq:E-2}) yield the same result
\begin{equation}
  E = q^2 \left(N + \case{3}{2} \pm \case{1}{2}\right) \left(N + 2k + \case{1}{2} \pm
  \case{1}{2}\right).  \label{eq:E-12}
\end{equation}
Comparison with (\ref{eq:E}) shows that only the upper sign choice in (\ref{eq:E-12}) leads to
physical wavefunctions $\Psi_{N, N-\nu}(x,y)$.\par
%
%---------------------------------------------------------------------------------------------
%
Restricting ourselves to such a choice, we can now rewrite all the results obtained in this subsection
in terms of $N$ and $\nu$ instead of $p$ and $m$. In particular, the two expressions
(\ref{eq:Phi-1}) and (\ref{eq:Phi-2}) for the structure function can be recast in a single form
$\Phi(m) \to \Phi_{\nu}$, where
\begin{equation}
  \Phi_{\nu} = 3 \cdot 2^{30} q^{20} \nu (\nu-1) (\nu+2k-1) (\nu+2k-2) (N+\nu+2k) (N+\nu+2k
  +1) (N-\nu+2) (N-\nu+3).  \label{eq:Phi-nu}
\end{equation}
\par
%
%----------------------------------------------------------------------------------------------
%
More importantly, our quadratic algebraic analysis provides us with an entirely new result, namely
the matrix elements of the integral of motion $L$ in the basis wherein $H$ and $R$ are
simultaneously diagonal. On using indeed the correspondence $|p, m\rangle \to \Psi_{N, N-\nu}$,
as well as the results in Eqs.~(\ref{eq:Fock}), (\ref{eq:B-para}), (\ref{eq:sigma}), (\ref{eq:rho})
and (\ref{eq:Phi-nu}), we obtain
\begin{equation}
  L \Psi_{N, N-\nu} = \sigma_{\nu} \Psi_{N, N-\nu} + \tau_{\nu} \Psi_{N, N-\nu+2} +
  \tau_{\nu+2} \Psi_{N, N-\nu-2},  \label{eq:L-me} 
\end{equation}
where we have reset $\sigma(m) \to \sigma_{\nu}$, $\rho(m) \to \rho_{\nu}$ and defined
$\tau_{\nu} = s_{\nu} \rho_{\nu-2} \sqrt{\Phi_{\nu}}$. The explicit form of the coefficients
 on the right-hand side of (\ref{eq:L-me}) is given by
\begin{eqnarray}
  \sigma_{\nu} & = & \frac{q^2}{2(\nu+k-1)(\nu+k+1)} \{- (\nu+k-1)^2 (\nu+k+1)^2 \nonumber
       \\ 
  && \mbox{}+ [N^2 + (2k+3)N + 2k^2 + 2k +1] (\nu+k-1) (\nu+k+1) \nonumber \\
  && \mbox{} - k(k-1)(N+k+1)(N+k+2)\}, \\
  \tau_{\nu}^2 & = & \frac{q^4}{16(\nu+k-2) (\nu+k-1)^2 (\nu+k)} \nu (\nu-1) (\nu+2k-1)
       (\nu+2k-2) \nonumber \\
  && \mbox{} \times (N-\nu+2) (N-\nu+3) (N+\nu+2k) (N+\nu+2k+1).  \label{eq:tau}
\end{eqnarray} 
Note that $\tau_{\nu}$ is determined up to some phase factor $s_{\nu}$ depending on the
convention adopted for the relative phases of $\Psi_{N, N-\nu}$ and $\Psi_{N, N-\nu + 2}$.\par
%
%-----------------------------------------------------------------------------------------------
%
{}For $N=4$, for instance, $\nu$ runs over 0, 2, 4, so that Eqs.~(\ref{eq:L-me}) -- (\ref{eq:tau})
become
\begin{eqnarray}
  L \Psi_{4,0} & = & \frac{q^2}{k+3} \Biggl[(13k+21) \Psi_{4,0} + 3s_4 \sqrt{\frac{2(k+1)(2k+3)
       (2k+9)}{k+2}} \Psi_{4,2}\Biggr], \\
  L \Psi_{4,2} & = & q^2 \Biggl[\frac{3s_4}{k+3} \sqrt{\frac{2(k+1)(2k+3)(2k+9)}{k+2}}
       \Psi_{4,0} + \frac{17k^2+76k+39}{(k+1)(k+3)} \Psi_{4,2} \nonumber \\
  && \mbox{} + \frac{s_2}{k+1} \sqrt{\frac{10(k+3)(2k+1)(2k+7)}{k+2}} \Psi_{4,4}\Biggr], \\
  L \Psi_{4,4} & = & \frac{q^2}{k+1} \Biggl[s_2 \sqrt{\frac{10(k+3)(2k+1)(2k+7)}{k+2}}
       \Psi_{4,2} + 5(k+3) \Psi_{4,4}\Biggr]. 
\end{eqnarray}
As a check, these results can be compared with those derived from the action of $L$ on the
expansions of $\Psi_{4,0}$, $\Psi_{4,2}$ and $\Psi_{4,4}$ in terms of the first basis
eigenfunctions $\psi_{0,4}$, $\psi_{1,2}$ and $\psi_{2,0}$ (see, e.g., Eqs.~(I3.61) and (I3.49) for
$\Psi_{4,0}$ and $\Psi_{4,4}$, respectively). This leads to the phase factors $s_2 = s_4 = -1$.\par
%
%--------------------------------------------------------------------------------------------------
% 
To conclude, it is worth mentioning that had we made the opposite choice in Eq.~(\ref{eq:A-B}),
i.e., $A=L$ and $B=R$, we would not have been able to use the deformed parafermionic realization
(\ref{eq:A-para}), (\ref{eq:B-para}) to determine the energy spectrum. The counterpart of the
parafermionic vacuum state would indeed have been a function annihilated by $L$ and therefore
constructed from the unphysical function $\bar{\chi}_{-1}(y)$ of Eq.~(\ref{eq:chi-bar}).\par
%
%=================================================================
% 
\section{Exactly solvable PDM models in three dimensions}

\setcounter{equation}{0}

In the present section, we plan to show that the Hamiltonian (\ref{eq:H2}) on the two-dimensional
domain (\ref{eq:D2}) can be easily extended to three dimensions in such a way that the domain
keeps its essential characteristic of abrupt termination while the Hamiltonian remains exactly
solvable. The latter will still be integrable with three independent integrals of motion $H$, $L$ and
$M$, but the superintegrability of the two-dimensional model will be lost. This generalization can be
carried out in two different ways.\par
%
%+++++++++++++++++++++++++++++++++++++++++++++++++++++
%
\subsection{Exactly solvable PDM model in a semi-infinite parallelepipedal channel}

In (\ref{eq:H2}), let us replace the operator $\partial^2_y$ by the two-dimensional Laplacian in
cartesian coordinates $\partial^2_y + \partial^2_z$ and assume that $z$ varies in the same range
as $y$. This leads us to the Hamiltonian
\begin{equation}
  H = - \partial_x \cosh^2 qx \partial_x - \cosh^2 qx (\partial^2_y + \partial^2_z) - q^2 \cosh^2
  qx + q^2 k(k-1) \csch^2 qx,
\end{equation}
defined on the semi-infinite parallelepipedal domain
\begin{equation}
  D: \qquad 0 < x < \infty, \qquad - \frac{\pi}{2q} < y, z < \frac{\pi}{2q},   
\end{equation}
with wavefunctions satisfying the conditions
\begin{equation}
  \psi(0,y,z) = 0, \qquad \psi\left(x, \pm \frac{\pi}{2q},z\right) = 0, \qquad \psi\left(x, y, \pm
  \frac{\pi}{2q}\right) = 0.  \label{eq:boundary3-1}  
\end{equation}
\par
%
%------------------------------------------------------------------------------------
%
Such a Hamiltonian commutes with the operators
\begin{equation}
  L = - \partial^2_y, \qquad M = - \partial^2_z.
\end{equation}
Their simultaneous normalizable eigenfunctions $\psi_{n,l,m}(x,y,z)$, fulfilling
(\ref{eq:boundary3-1}), can be easily obtained along the lines detailed in Section 3.1 of I. They can
be written as
\begin{equation}
  \psi_{n,l,m}(x,y,z) = \phi_{n,l,m}(x) \chi_l(y) \zeta_m(z),
\end{equation}
where $\chi_l(y)$ is given in (\ref{eq:chi}), $\zeta_m(z)$ can be expressed in a similar way with
$m$ and $z$ substituted for $l$ and $y$, respectively, while
\begin{equation}
  \phi_{n,l,m}(x) = {\cal N}_{n,l,m} (\tanh qx)^k (\sech qx)^{1+\delta}
       P^{\left(k-\case{1}{2}, \delta\right)}_n(1 - 2 \tanh^2 qx),  \label{eq:phi3}
\end{equation}
with
\begin{eqnarray}
  \delta & = & \sqrt{(l+1)^2 + (m+1)^2}, \\
  {\cal N}_{n,l,m} & = & \left(\frac{2q \left(2n + k + \frac{1}{2} + \delta\right) n!\, \Gamma\left(n
       + k + \frac{1}{2} + \delta\right)}{\Gamma(n+1+\delta) \Gamma\left(n + k +
       \frac{1}{2}\right)}\right)^{1/2}.
\end{eqnarray}
\par
%
%---------------------------------------------------------------------------------------
%
The simultaneous eigenvalues of $L$, $M$ and $H$ are $(l+1)^2 q^2$, $(m+1)^2 q^2$ and
\begin{equation}
  E_{n,l,m} = q^2 (2n+1+\delta) (2n+2k+\delta),  \label{eq:E3}
\end{equation}
where $n$, $l$, $m=0$, 1, 2,~\ldots. The only remaining degeneracies are those connected with
the $(l, m)$ exchange, i.e., $E_{n,l,m} = E_{n,m,l}$ for $l \ne m$, as well as some `accidental'
degeneracies, such as $E_{n,1,8} = E_{n,5,6}$ corresponding to $\delta = \sqrt{85}$.\par
%
%+++++++++++++++++++++++++++++++++++++++++++++++++++++++++
%
\subsection{Exactly solvable PDM model in a semi-infinite cylindrical channel}

Alternatively, we may replace $\partial^2_y$ in (\ref{eq:H2}) by the two-dimensional Laplacian in
polar coordinates $\partial^2_{\rho} + \frac{1}{\rho} \partial_{\rho} + \frac{1}{\rho^2}
\partial^2_{\varphi}$ and assume that $\rho$ varies in a finite domain, on the boundary of which
wavefunctions vanish. In this way, we get the Hamiltonian
\begin{equation}
  H = - \partial_x \cosh^2 qx \partial_x - \cosh^2 qx \left(\partial^2_{\rho} + \frac{1}{\rho} 
  \partial_{\rho} + \frac{1}{\rho^2} \partial^2_{\varphi}\right) - q^2 \cosh^2
  qx + q^2 k(k-1) \csch^2 qx,
\end{equation}
defined on the semi-infinite cylindrical domain
\begin{equation}
  D: \qquad 0 < x < \infty, \qquad 0 \le \rho < R, \qquad 0 \le \varphi < 2\pi,   
\end{equation}
with wavefunctions such that
\begin{equation}
  \psi(0,\rho,\varphi) = 0, \qquad \psi(x,R,\varphi) = 0,  \qquad \psi(x,\rho,2\pi) = \psi(x,\rho,0). 
  \label{eq:boundary3-2}  
\end{equation}
\par
%
%------------------------------------------------------------------------------------
%
The two operators commuting with $H$ are now
\begin{equation}
  L = - \left(\partial^2_{\rho} + \frac{1}{\rho} \partial_{\rho} + \frac{1}{\rho^2}
  \partial^2_{\varphi}\right), \qquad M = - {\rm i} \partial_{\varphi}.
\end{equation}
The simultaneous normalizable eigenfunctions of $H$, $L$ and $M$ can be written as
\begin{equation}
  \psi_{n,m,s}(x,\rho,\varphi) = \phi_{n,|m|,s}(x) \chi_{|m|,s}(\rho) \zeta_m(\varphi).
\end{equation}
Here
\begin{equation}
  \zeta_m(\varphi) = \frac{1}{\sqrt{2\pi}} e^{{\rm i} m\varphi}
\end{equation}
corresponds to the eigenvalues $m=0, \pm 1, \pm 2, \ldots$ of $M$. Furthermore,
\begin{equation}
  \chi_{|m|,s}(\rho) = {\cal N}_{|m|,s} J_{|m|}(\kappa_{|m|,s} \rho), \qquad \kappa_{|m|,s} =
  \frac{j_{|m|,s}}{R},
\end{equation}
where $J_{|m|}(z)$ is a Bessel function, the symbol $j_{|m|,s}$, $s=1$, 2,~\ldots, conventionally
denotes~\cite{abramowitz} its real, positive zeros, and~\cite{gradshteyn}
\begin{equation}
  {\cal N}_{|m|,s} = \sqrt{2} \left[R J_{|m|+1}(j_{|m|,s})\right]^{-1},
\end{equation}
provides normalized solutions to the eigenvalue equation
\begin{equation}
  L \chi_{|m|,s}(\rho) \zeta_m(\varphi) = \kappa_{|m|,s}^2 \chi_{|m|,s}(\rho) \zeta_m(\varphi), 
\end{equation}
which satisfy the second condition in (\ref{eq:boundary3-2}).\footnote{For the solution of a
similar problem see \cite{flugge}.} Finally, $\phi_{n,|m|,s}(x)$ and the energy eigenvalues
$E_{n,|m|,s}$ are still given by the right-hand sides of Eqs.~(\ref{eq:phi3}) and (\ref{eq:E3}), but 
with $\delta$ now defined by
\begin{equation}
  \delta = \frac{\kappa_{|m|,s}}{q} = \frac{j_{|m|,s}}{qR}.
\end{equation}
This time the only level degeneracy left is that connected with the sign of $m$.\par
%
%==========================================================
% 
\section{Conclusion}

In this paper, we have revisited the exactly solvable PDM model in a two-dimensional semi-infinite
layer introduced in I. Here we have taken advantage of its superintegrability with two integrals of
motion $L$ and $R$ that are quadratic in the momenta to propose a third method of solution in the
line of some recent analyses of such problems.\par
%
%----------------------------------------------------------------------------------------
%
We have first determined the explicit form of the quadratic associative algebra generated by $L$,
$R$ and their commutator. We have shown that it is a quadratic Racah algebra QR(3) and that its
Casimir operator $K$ is a second-degree polynomial in $H$. We have also obtained the quadratic
Poisson algebra arising in the classical limit.\par
%
%--------------------------------------------------------------------------------------
%
We have then studied the finite-dimensional irreducible representations of our algebra in a basis
wherein $K$ (or $H$) and $R$ are diagonal. For such a purpose, we have used a simple procedure,
proposed in \cite{daska01}, consisting in mapping this basis onto deformed parafermionic oscillator
states of order $p$. Among the results so obtained for the energy spectrum, we have selected
those with which physical wavefunctions can be associated. This has illustrated once again the
well-known fact that in quantum mechanics the physics is determined not only by algebraic
properties of operators, but also by the boundary conditions imposed on wavefunctions. Our
analysis has provided us with an interesting new result, not obtainable in general form in the
SUSYQM approach of I, namely the matrix elements of $L$ in the basis wherein $H$ and $R$ are
simultaneously diagonal.\par
%
%-----------------------------------------------------------------------------------------------
%
In the last part of our paper, we have extended our two-dimensional model to three dimensions in
two different ways by considering either a semi-infinite parallelepipedal channel or a semi-infinite
cylindrical one. Both resulting models remain integrable and exactly solvable, but the
superintegrability of the two-dimensional model is lost. From a physical viewpoint, they illustrate
the generation of bound states in a quantum channel when the uniformity is broken by an abrupt
termination.\par
%
%--------------------------------------------------------------------------------------------
%
As a final point, it is worth observing that the procedure used here to construct the irreducible
representations of QR(3) is not the only one available. In particular, the ladder-operator method
employed in~\cite{granovskii92a, granovskii92b} would allow us to express the transformation
matrix elements between the bases $\psi^{(k)}_{n,l}$ and $\Psi^{(k)}_{N,N_0}$ (denoted by
$Z^{(k)}_{N_0;n,l}$ in I) in terms of Racah-Wilson polynomials.\par
%
%========================================================
%

\end{document}